\documentstyle[aps,epsf,preprint]{revtex}
\tightenlines
\begin{document}
\draft
\title{Contribution to muon g-2 from the \boldmath{$\pi^0\gamma$} and \boldmath{$\eta\gamma$} intermediate states in
 the vacuum polarization}
\author
{N.N. Achasov
\thanks{achasov@math.nsc.ru}
\ and A.V. Kiselev
\thanks{kiselev@math.nsc.ru}}

\address{
   Laboratory of Theoretical Physics,
 Sobolev Institute for Mathematics, Novosibirsk, 630090,
Russia}
\date{\today}
\maketitle

\begin{abstract}
Using new experimental data, we have calculated the contribution to the anomalous magnetic moment of the muon from the
  $\pi^0\gamma$ and $\eta\gamma$ intermediate states in the vacuum polarization with high precision:
   $a_{\mu}(\pi^0\gamma)+a_{\mu}(\eta\gamma)=(54.7\pm 1.5)\times 10^{-11}$. We have also found the small contribution
    from $e^+e^-\pi^0,\ e^+e^-\eta$ and $\mu^+\mu^-\pi^0$ intermediate states equal to $0.5\times 10^{-11}$.

\end{abstract}


New experimental data \cite{snd,snd2,cmd} allows to calculate contribution to the anomalous magnetic moment of the muon $a_{\mu}\equiv\frac{g_{\mu}-2}{2}$ from the $\pi^0\gamma$ and $\eta\gamma$
intermediate states in the vacuum polarization with high precision. We have also found the contribution from $e^+e^-\pi^0,\ e^+e^-\eta$ and $\mu^+\mu^-\pi^0$ intermediate states.

The contribution to $a_{\mu}$ from the arbitrary intermediate state X ( hadrons, hadrons+$\gamma$, etc.) in the vacuum polarization can be obtained via the dispersion integral
\begin{equation}
a_{\mu}=\Big(\frac{\alpha m_{\mu}}{3\pi}\Big)^2\int\frac{ds}{s^2}K(s)R(s).
\end{equation}

$$R(s)\equiv\frac{\sigma(e^+e^-\rightarrow X)}{\sigma(e^+e^-
\rightarrow \mu^+\mu^-)},\hspace{3mm} \sigma(e^+e^-\rightarrow \mu^+\mu^-)\equiv \frac{4\pi\alpha^2}{3s}.$$

$$K(s>4m_{\mu}^2)=\frac{3s}{m_{\mu}^2}\Big\{x^{2}(1-\frac{x^2}{2})+(1+x)^2(1+\frac{1}{x^2})\Big[\ln(1+x)-x+
\frac{x^2}{2}\Big]+\frac{1+x}{1-x}x^2 \ln(x)\Big\}=$$

$$=\frac{3}{a^3}\Big(16(a-2)\ln\frac{a}{4}-2a(8-a)-8(a^2-8a+8)
\frac{\mbox{arctanh}(\sqrt{1-a})}{\sqrt{1-a}}\Big),$$
$$x=\frac{1-\sqrt{1-\frac{4m{_\mu }^2}{s}}}{1+\sqrt{1-\frac{4m{_\mu }^2}
{s}}},
\hspace{5mm} a=\frac{4m_{\mu}^2}{s}.$$

$$K(s<4m_{\mu}^2)=\frac{3}{a^3}\Big(16(a-2)\ln\frac{a}{4}-2a(8-a)-8(a^2-8a+8)
\frac{\arctan(\sqrt{a-1})}{\sqrt{a-1}}\Big).$$

Evaluating integral (1) with the trapezoidal rule for the experimental data from SND \cite{snd,snd2}, see Fig. 1(a), we found the contribution of $\pi^0\gamma$:
\begin{equation}
a_{\mu}(\pi^0\gamma)=(46.2\pm.6\pm1.3)\times10^{-11} , \hspace{5mm}
600\ MeV<\sqrt{s}<1039\ MeV.
\end{equation}

The first error is statistical, the second is systematic.
For the energy region $\sqrt{s}<600$ MeV we used theoretical formula for
the cross-section:
\begin{equation}
\sigma(e^+e^-\rightarrow\pi^0\gamma)=\frac{8\alpha f^2}{3}\Big(1-
\frac{m_{\pi^0}^2}{s}\Big)^3\frac{1}{\Big(1-\frac{s}{m_{\omega}^2}\Big)^2},
\end{equation}
where $f^2=\frac{\pi}{m_{\pi^0}^3}\Gamma_{\pi^0\rightarrow\gamma\gamma}\cong
10^{-11}/MeV^2$ according to \cite{pdg}. Eq. (3) has been written in the approximation
\begin{equation}
\Gamma_{\rho}=\Gamma_{\omega}=0,\hspace{3 mm} m_{\rho}-m_{\omega}=0.
\end{equation}
The $\gamma*\rightarrow \pi^0\gamma$ amplitude is normalized on the $\pi^0\rightarrow\gamma\gamma$ one at $s=0$. The result is
\begin{equation}
a_{\mu}(\pi^0\gamma)=1.3\times10^{-11} , \hspace{5mm} \sqrt{s}<600\ MeV.
\end{equation}

Note that the region $\sqrt{s}<2m_{\mu}$ gives the negligible contribution $2\times 10^{-13}$.

We neglect the small errors dealing with the experimental error in the width $\Gamma_{\pi^0\rightarrow\gamma\gamma}$ (7\%) and the approximation (4) (1.5\%).

The Eq. (3) agrees with the data in the energy region $\sqrt{s}<700$ MeV, at higher energies the approximation (4) does not work carefully, see Fig. 1(b).

If we use the point-like model, as in \cite{ynd}, we will get Eq. (3) without factor $\Big(1-\frac{s}{m_{\omega}^2}\Big)^{-2}$. This formula predicts the contribution from low energies several times less than (5), see also Fig. 1(b).

Treating the data from CMD-2 \cite{cmd} in the same way, we get contribution of $\eta\gamma$:
\begin{equation}
a_{\mu}(\eta\gamma)=(7.1\pm .2\pm .3)\times10^{-11} , \hspace{5mm} 720\
MeV<\sqrt{s}<1040\ MeV. \end{equation}

According to the quark model (and the model of vector dominance also), the energy region $\sqrt{s}<720$ MeV is dominated by the $\rho$-resonance, hence $\sigma(e^+e^-\rightarrow\eta\gamma)\cong\sigma(e^+e^-\rightarrow\rho\rightarrow\eta\gamma)$.
So we change Eq. (3) according to this fact, take into account the $\rho$ width and get the small contribution:
\begin{equation}
a_{\mu}(\eta\gamma)=.1\times 10^{-11},\hspace{5mm} \sqrt{s}<720\ MeV,
\end{equation}

Summing (2), (5), (6) and (7), we can write
\begin{equation}
a_{\mu}(\pi^0\gamma)+a_{\mu}(\eta\gamma)=(54.7\pm .6\pm 1.4)\times 10^{-11},
\end{equation}
where statistical and systematic errors are separately added in
quadrature. In Table 1 we present our results with statistical and
systematic errors added in quadrature. Comparing Eq. (8) with the
analogous calculation in \cite{ynd} (see Table 1), one can see
that our result is 27\% more and the error is 2.5 times less. The
contribution (8) accounts for 1.37 of the projected error of the
E821 experiment at Brookhaven National Laboratory ($40\times
10^{-11}$) or 36\% of the reached accuracy ($150\times 10^{-11}$
\cite{bnl}).

We can also take into account the intermediate state $\pi^0e^+e^-$, using the obvious relation

\begin{equation}
\sigma (e^+e^-\rightarrow\pi^0 e^+e^-,s)=\frac{2}{\pi}\int_{2m_e}^{\sqrt{s}-m_{\pi^0}} \frac{dm}{m^2}\Gamma_{\gamma *\to e^+e^-}(m)\sigma(e^+e^-\rightarrow\pi^0\gamma *,s,m),
\end{equation}
where m is the invariant mass of the $e^+e^-$ system, $\Gamma_{\gamma *\to e^+e^-}(m)$=$(1/2)\alpha \beta_e m(1-\beta_e^2/3),\ \beta_e$=$\sqrt{1-4m_{e}^2/m^2},\ \sigma(e^+e^-\rightarrow\pi^0\gamma *,s,m)=\Big(p(m)/p(0)\Big)^3\sigma(e^+e^-\rightarrow\pi^0\gamma,s),\ p(m)=(\sqrt{s}/2)\sqrt{(1-(m_{\pi^0}+m)^2/s)(1-(m_{\pi^0}-m)^2/s)}$ is the momentum of $\gamma *$ in s.c.m.

In the same way we can calculate $a_{\mu}(\mu^+\mu^-\pi^0)$ and $a_{\mu}(e^+e^-\eta)$. The result is

\begin{equation}
a_{\mu}(e^+e^-\pi^0)+a_{\mu}(\mu^+\mu^-\pi^0)+a_{\mu}(e^+e^-\eta)=(.4+.026+.057)\times 10^{-11}=.5\times 10^{-11}.
\end{equation}

Note that if $m\agt m_{\rho}$ we have the effect of the excitation of resonances in the reaction $e^+e^-\to\pi^0(\rho,\ \omega)\to\pi^0e^+e^-$. However this effect increases the final result (10) less than by 10\% because of the factor $(p(m)/p(0))^3$, which suppresses the high m. So we ignore this correction. We also neglect $a_{\mu}(\mu^+\mu^-\eta)=2\times 10^{-14}$.

As it was noted in \cite{ynd} and \cite{jeg}, it is necessary to take into account also
$$a_{\mu}(hadrons +\gamma,rest)=a_{\mu}(\pi^+\pi^-\gamma)+a_{\mu}(\pi^0\pi^0\gamma)
+a_{\mu}(hadrons+\gamma,\ s>1.2\ GeV^2).$$

We take $a_{\mu}(\pi^+\pi^-\gamma)=(38.6\pm 1.0)\times 10^{-11}$ from \cite{jeg} (see also \cite{ynd}), $a_{\mu}(\pi^0\pi^0\gamma)
+a_{\mu}(hadrons+\gamma,\ s>1.2\ GeV^2)=(4\pm 1)\times 10^{-11}$ from \cite{ynd}. Adding this to (8), we get

\begin{equation}
a_{\mu}(hadrons +\gamma,\ total)=(97.3\pm 2.1)\times 10^{-11}.
\end{equation}

The contribution (11) accounts for 2.43 of the projected error of the E821 experiment or 65\% of the reached accuracy.

In fact, the errors in (8) and (11) are negligible for any imaginable $(g-2)_{\mu}$ measurement in near future.

\begin{center}
\begin{tabular}{|c|c|c|}
\multicolumn{3}{c}{Table 1. Contribution to $a_{\mu}\times
10^{11}$} \\ \hline

State  & Our value  & Ref. \cite{ynd} \\ \hline $\pi^0\gamma$ &
$\hspace{1mm} 47.5\pm 1.4 \hspace{1mm}$ & $37\pm 3$ \\ \hline
$\eta\gamma$ & $7.2\pm .4$ & $ \hspace{1mm} 6.1\pm 1.4
\hspace{1mm}$ \\ \hline $\pi^0\gamma+\eta\gamma$ & $54.7 \pm 1.5 $
& $43\pm 4$ \\ \hline hadrons+$\gamma$, total & $97.3\pm 2.1$ &
$93\pm 11$ \\ \hline
\end{tabular}
\end{center}

\begin{figure}
\centerline{
\epsfxsize=8 cm \epsfysize=8cm \epsfbox{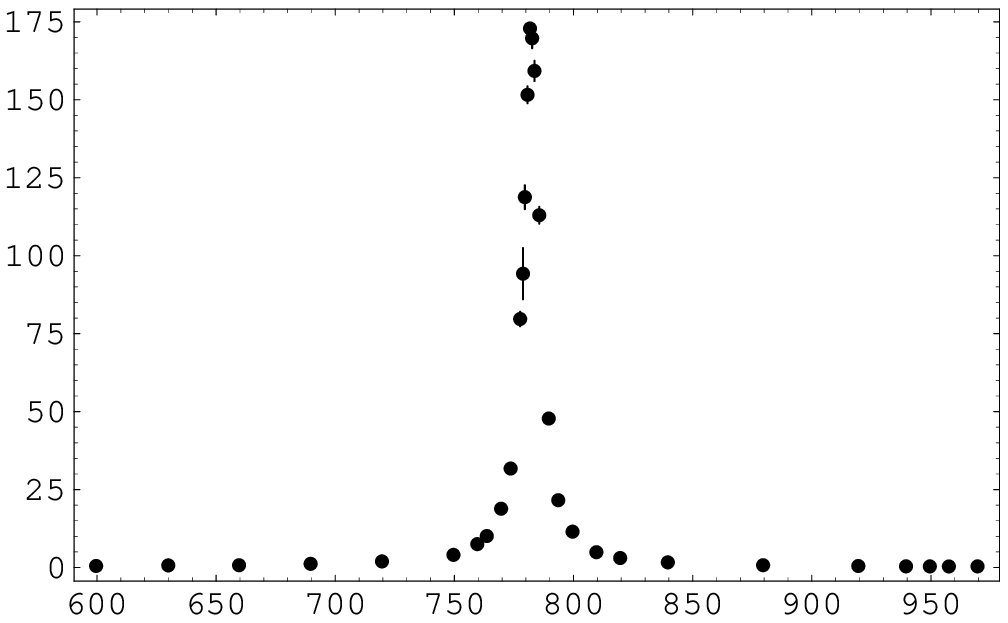}
\hfill
\epsfxsize=8 cm \epsfysize=8cm \epsfbox{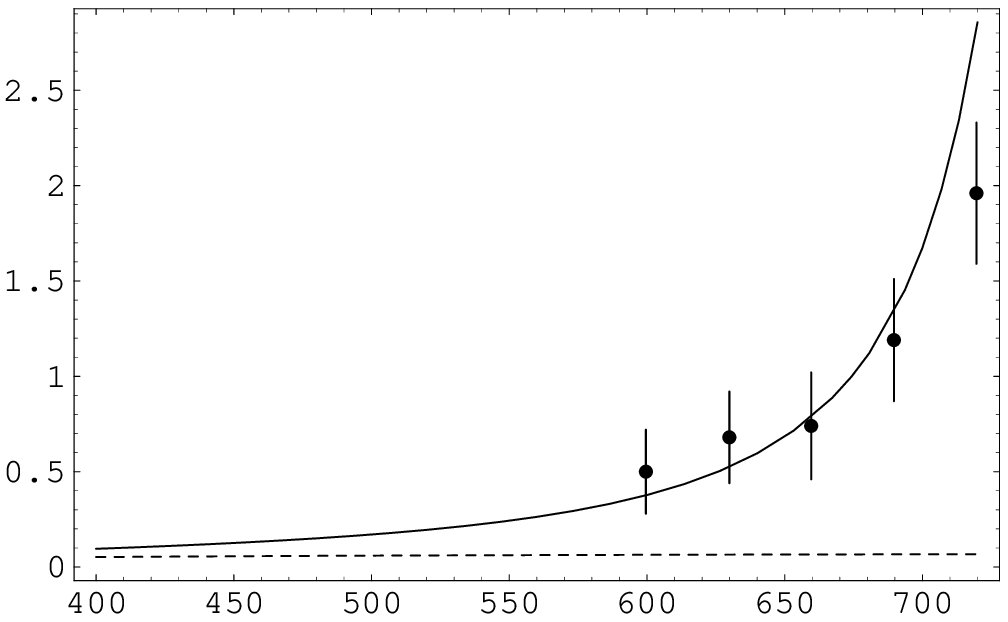}
}
\parbox[t]{\textwidth}{\hspace{3.8cm} a) \hspace{7.5cm} b)}
\\
 \caption{ a) Plot of the dependence $\sigma(e^+e^-\to\pi^0\gamma)$, nb upon $\sqrt{s}$, MeV (SND experimental data).
 b) Comparison of the theoretical formulas for $\sigma(e^+e^-\rightarrow\pi^0\gamma)$. Eq. (3) is shown with the solid line, point-like model prediction is shown with the dashed line.
  } \label{fig1}
\end{figure}

\end{document}